# Time-resolved laser speckle contrast imaging (TR-LSCI) of cerebral blood flow

Faraneh Fathi,[1] Siavash Mazdeyasna,[1] Dara Singh,[1] Chong Huang,[1] Mehrana Mohtasebi,[1] Xuhui Liu,[1] Samaneh Rabienia Haratbar,[1] Mingjun Zhao,[1] Li Chen,[2] Arin Can Ulku,[3] Paul Mos,[3] Claudio Bruschini,[3] Edoardo Charbon,[3] Lei Chen,[4] Guoqiang Yu,[1,*]

[1]*Department of Biomedical Engineering, University of Kentucky, Lexington, Kentucky, 40506*
[2]*Biostatistics and Bioinformatics Shared Resource Facility, Markey Cancer Center, University of Kentucky, Lexington, Kentucky, 40536*
[3]*School of Engineering, Ecole polytechnique fédérale de Lausanne, Neuchâtel 2002, Switzerland*
[4]*Department of Physiology, Spinal Cord and Brain Injury Research Center, University of Kentucky, Lexington, Kentucky, 40506*
*\*Corresponding author.*
*E-mail Address: guoqiang.yu@uky.edu (G. Yu).*

**Abstract:** To address many of the deficiencies in optical neuroimaging technologies such as poor spatial resolution, time-consuming reconstruction, low penetration depth, and contact-based measurement, a novel, noncontact, time-resolved laser speckle contrast imaging (TR-LSCI) technique has been developed for continuous, fast, and high-resolution 2D mapping of cerebral blood flow (CBF) at different depths of the head. TR-LSCI illuminates the head with picosecond-pulsed, coherent, widefield near-infrared light and synchronizes a newly developed, high-resolution, gated single-photon avalanche diode camera (SwissSPAD2) to capture CBF maps at different depths. By selectively collecting diffuse photons with longer pathlengths through the head, TR-LSCI reduces partial volume artifacts from the overlying tissues, thus improving the accuracy of CBF measurement in the deep brain. CBF map reconstruction was dramatically expedited by incorporating highly parallelized computation. The performance of TR-LSCI was evaluated using head-simulating phantoms with known properties and *in-vivo* rodents with varied hemodynamic challenges to the brain. Results from these pilot studies demonstrated that TR-LSCI enabled mapping CBF variations at different depths with a sampling rate of up to 1 Hz and spatial resolutions ranging from tens of micrometers on the head surface to 1-2 millimeters in the deep brain. With additional improvements and validation in larger populations against established methods, we anticipate offering a noncontact, fast, high-resolution, portable, and affordable brain imager for fundamental neuroscience research in animals and for translational studies in humans.

## 1. Introduction

Cerebral blood flow (CBF) is vital for delivery of nutrients and oxygen to brain tissues and clearance of metabolic wastes [1, 2]. Abnormal CBF (hypo-perfusion or hyper-perfusion) contributes to many pathological conditions, including cerebral hypoxic ischemia, traumatic brain injury, ischemic stroke, malignant hypertension, and cerebral hemorrhage [3-7]. These pathological conditions are often associated with disruptions in CBF autoregulation and neurovascular coupling [8]. To comprehend the underlying causes and develop medical therapies for neurological and cerebral diseases, continuous and longitudinal monitoring of CBF fluctuations using noninvasive or minimally invasive imaging modalities is essential [9].

Currently available technologies for CBF measurements include dynamic susceptibility magnetic resonance imaging (MRI), positron emission tomography (PET), x-ray computed tomography (CT), and transcranial doppler ultrasound (TCD). While useful, these imaging modalities present a host of issues that restrict their applications [10-12]. MRI, PET, and CT are large, expensive, and difficult to use for continuous and longitudinal monitoring of brain hemodynamics [13]. PET uses ionizing radiation and CT involves exposing the patient to a significant amount of radiation, thus limiting their clinical applications [11]. TCD enables the detection of blood flow in large vessels but is not sensitive to microcirculation [14].

By contrast, optical imaging instruments are portable, inexpensive, and fast, enabling continuous assessment of cerebral hemodynamics in the microvasculature at the bedside [15-24]. Laser speckle contrast imaging (LSCI) and optical intrinsic signal imaging (OISI) with widefield illumination and charge-coupled device/complementary-metal-oxide-semiconductor (CCD/CMOS) camera detection enable fast and high-resolution 2D mapping of CBF and cerebral oxygenation on cortices of rodents, respectively [16, 21, 24, 25]. However, due to their limited penetration depth (<1 mm), LSCI and OISI require the retraction of the rodent's scalp and/or an invasive cranial window (especially for rats with thicker skulls) for cortical imaging. As a result, LSCI and OISI are difficult to use for noninvasive deep brain imaging.

Conventional diffuse optical technologies such as near-infrared spectroscopy (NIRS) and diffuse correlation spectroscopy (DCS) use continuous-wave near-infrared point sources and discrete photodetectors to noninvasively measure cerebral blood oxygen saturation ($ScO_2$) and CBF respectively in deep brains (up to centimeters) [26-31]. Correspondingly, near-infrared diffuse optical tomography (DOT) and diffuse correlation tomography (DCT) use dense arrays of sources and detectors to yield boundary measurements across numerous source-detector (S-D) pairs and solve inverse problems to reconstruct 3D images of $ScO_2$ and CBF, respectively [32-37]. Tomographic 3D reconstructions by these methods need offline solving of ill-posed nonlinear inverse problems, which are complex and very time consuming [38]. More recently, time-resolved NIRS (TR-NIRS) and time-resolved DCS (TR-DCS) use pulsed point sources and discrete single-photon avalanche diodes to measure temporal point-spread functions and autocorrelation functions for quantifying $ScO_2$ and CBF in deep brains, respectively [39-45]. However, most systems suffer from limited numbers of discrete sources and photodetectors, thus taking sparse S-D pair measurements on a limited region-of-interest (ROI) of the head. Expanding the ROI to cover a significant portion of the head introduces great challenges in high-channel-count instrumentation, imaging array ergonomics, and data quality management. As a result, their spatial resolutions [46] and head coverages (ROIs) are limited, thus inhibiting the localization of regional brain activations [47]. Moreover, collected cerebral signals are inherently influenced by partial volume artifacts from overlayer tissues of scalp and skull [48, 49].

To address many of the deficiencies in competing neuroimaging technologies, we are developing a novel, noncontact, affordable, time-resolved laser speckle contrast imaging (TR-LSCI) technique that enables continuous, fast, and high-resolution 2D mapping of CBF at different depths over a large ROI on the head [50]. In contrast to other near-infrared technologies using point-source illuminations and discrete detectors, TR-LSCI illuminates picosecond-pulsed, coherent, widefield near-infrared light onto the head and synchronizes a newly developed gated single-photon avalanche diode (SPAD) camera to rapidly image CBF distributions at different depths of the brain. By applying the time-gated strategy to synchronize the pulsed widefield illumination and camera opening time (i.e., time-resolved method), TR-LSCI differentiates short and long photon paths through layered head tissues (i.e., scalp, skull, and brain) to precisely map CBF distributions in subjects with different head scales (i.e., multiscale). Importantly, TR-LSCI eliminates the need for time-consuming complex tomographic reconstruction of 3D CBF images in conventional DOT/DCT technologies, thereby offering depth-sensitive 2D CBF maps in near real time.

After assembling a benchtop prototype, we carried out proof-of-concept studies to introduce the innovative TR-LSCI as a photon pathlength resolved technology for imaging of CBF variations at different depths of the head. Head/flow-simulating phantoms with different thicknesses of top layers were created and imaged to evaluate depth sensitivity of TR-LSCI technology. *In vivo* adult rodents were then imaged by the TR-LSCI during $CO_2$ inhalations (increasing CBF globally) and during unilateral and bilateral transient ligations of carotid arteries (reducing CBF regionally and globally). Results verified the capability of TR-LSCI for continuous, fast, and high-density, 2D mapping CBF changes at different depths.

## 2. Methods and Materials

### 2.1 TR-LSCI Prototype

**Fig. 1** shows our benchtop TR-LSCI device setting on an optical table. A free space, picosecond-pulsed, single-mode laser worked as a coherence point source (wavelength: 775 nm; pulse width: 540 ps; spectral bandwidth: <1 nm; max power at 20 MHz: 2 Watt; Katana-08 HP, NKT Photonics). Two engineered optical diffusers (ED1-C20-MD and ED1-S20-MD, Thorlabs) were placed in front of the point source to generate the widefield illumination. The laser power was adjusted so that the maximum photon counts detected by the gated SPAD camera (SwissSPAD2, EPFL, Switzerland) were within its linearity range (<500 counts for 10-bit image). Correspondingly, the power of incident light from the TR-LSCI on the tissue surface was less than 1.2 mW/cm$^2$ (measured by a power meter), which is compliant with the Accessible Emission Limit Class 3R of the American National Standards Institute (ANSI) standard [51].

SwissSPAD2 is an ultra-high-speed single-photon camera, comprising 512 × 512 SPAD pixels that can be time-gated [52-54]. In our experiments we used a resolution of 256 × 512 pixels to capture speckle contrast images at different depths. SwissSPAD2 achieves a combination of high sensitivity (50% photon detection probability at 520 nm), low pitch (16.38 µm), moderate fill factor (10.5% fill factor equivalent to 3 µm SPAD active area radius), high temporal resolution (18 ps minimum gate shift), and low dark count rate (0.26 photon counts per second (cps)/µm$^2$). Compared to the majority of other SPAD detectors, the combination of large spatial resolution (large number of pixels) and high temporal resolution (picosecond range) is the greatest advantage of the SwissSPAD2 camera [53].

A zoom lens (MLM3X-MP, Computar) was coupled to the camera for adjusting ROI size. The F/# of the zoom lens was set at 11 to satisfy the Nyquist criterion based on laser speckle and camera pixel sizes [55]. A pair of polarizers (LPNIRE050-B and LPNIRE200-B, Thorlabs) were added crossing the source and detection paths to reduce the influence of source reflections directly from the tissue surface. The room light was dimmed during experiments to minimize the impact of ambient light.

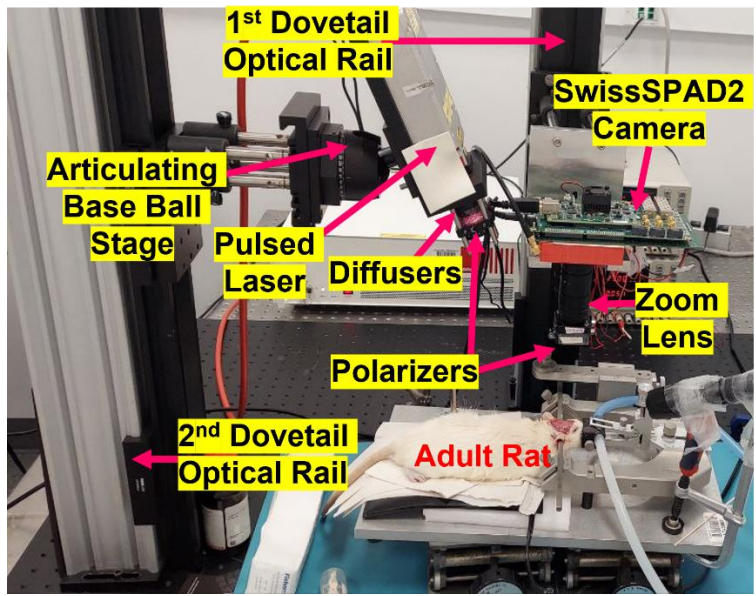

**Figure 1: TR-LSCI prototype for 2D mapping of CBF distributions at different depths in an adult rat.** The picosecond-pulsed laser and SwissSPAD2 camera used for the benchtop TR-LSCI prototype were supported by two dovetail optical rails fixed on an optical table. Side illumination setup was achieved by placing the laser on the side of the camera. The pulse laser produced triggers at 20 MHz to synchronize the laser and camera. The rat on the heating blanket was anesthetized (1-2% isoflurane) with its head secured on a stereotaxic frame. Noncontact TR-LCSI was placed over the head to continuously image CBF. Major components of the TR-LSCI device include a picosecond-pulsed laser to generate a coherent point light, two optical diffusers to generate a widefield illumination, a pair of polarizers to reduce source reflections, a zoom lens to adjust the ROI, and a SwissSPAD2 camera to capture speckle contrast images at different depths.

### *2.2 TR-LSCI Principle and Data Acquisition*

SwissSPAD2 can operate with two modes: intensity mode and gated mode. The intensity mode used as the conventional LSCI operates by having the gate fully open during the exposure window and fully closed during the readout window. In the intensity mode, images are captured with an 8-bit depth, whereas the gated mode allows for capturing images with both 8-bit and 10-bit depths. **Fig. 2** shows the principle and data acquisition of TR-LSCI with the gated mode. Time gating in SPAD cameras functions as a filter that only captures photons when they arrive within a specific time window [54] . Through electrical triggers produced by the pulsed laser as the master control, the laser and SwissSPAD2 camera are synchronized at 20 MHz (equivalent period of 50 ns = 1/20 MHz) for acquiring multiple gated images at different delay times (minimal 18 picoseconds). Overlapping gates are created when the delay time between adjacent gate windows (also known as gate step) is less than the actual camera gate width (13.1 ns used in this study). Using overlapping gates with a minimal gate delay time of 18 ps, sub-nanosecond temporal resolutions are achieved [56].

While the pulsed laser illuminates the tissue at 20 MHz, the SwissSPAD2 camera shifts its gated window every delay time (minimal 18 ps) over maximal 200 gate steps, thus collecting photons at different depths inside the target tissue volume (**Fig. 2a**). Specifically, after acquiring a 1st gated image at the shallowest depth, the gate shifts one delay time to the next step with respect to the reference from the laser pulse. Data collections are then repeated over up to 200 gates until the final gate position is reached (i.e., 200th gate at the deepest depth). For illustrative clarity in **Fig. 2a**, only 7 gate positions within a time interval of 10 ns (10 × 1 ns delay time) are illustrated instead of the full 200 gates.

Before data acquisition, it is crucial to determine the offset time to ensure that the 1$^{st}$ gated image is taken from the tissue surface (i.e., the shallowest depth). The offset is defined as the number of gates to be skipped before useful data acquisition (e.g., from tissue surface) begins. The offset time depends on TR-LSCI instrument configuration such as laser and camera synchronization time, light speed, and working distance. To determine the offset time, the SwissSPAD2 camera captured 200 gated images with a delay time of 1 ns over a total of 200 ns (**Fig. 2b**). Based on the gate intensity profile taken from a calibration tissue phantom, offsets between 30-32 ns (shaded region in **Fig. 2b**) were used in our TR-LSCI setup. For fair comparisons, all experiments in tissue phantoms and animals utilized the constant offset time of 30.5 ns.

After determining the offset time, data acquisition in our designed experiments started with the minimal delay time of 18 ps over 200 gates to maximize depth sensitivity of TR-LSCI measurements (**Fig. 2c**). As a result, the total detection time window over 200 gates was 3.6 ns (200 × 18 ps).

To collect one binary frame image at each gate step, the field-programmable gate array (FPGA) implementation in the SwissSPAD2 program involves opening the camera 400 times for data acquisition. The relation between the number of binary frames ($Y$) and the bit-depth ($b$) is defined as $Y = 2^b - 1$ [57-59]. However, since 10-bit images are constructed from four individually saved 8-bit images, $Y$ is equal to 1,020 in 10-bit mode ($b = 10$), instead of 1,023. Therefore, a total of 1020 binary frames are integrated in the FPGA to construct one 10-bit image at each gate step within ~20 ms (1020 × 400 × 50 ns). During this acquisition time (~20 ms), the total active exposure time for one 10-bit image is ~5.3 ms (1020 × 400 × 13.1 ns). As such, the total time to acquire and save one 10-bit image to a hard drive in the computer is ~30 ms (20 ms for data collection plus 10 ms for data readout and storage) [60].

The frame sampling time of 30 ms is equivalent to a frame rate of ~33 Hz. However, the actual highest frame rate in our experiments was only ~0.2 Hz. The actual SwissSPAD2 frame rates depend on multiple factors including binary frame readout, exposure time (acquisition time), and number of acquired gates. Since the SwissSPAD2 camera operates in global shutter mode, the sensor is insensitive to photons during read-out. Also, data acquisition must be stopped when image data are transferred from the RAM to the hard drive in the computer. Lastly, data acquisition is currently performed through 32-bit MATLAB code, which is not optimal for fast sampling and slower than 64-bit MATLAB. Future optimization of the SwissSPAD2 operating system with new editions of Python code or C++ would contribute to improving the frame rate. Additionally, incorporating a powerful computer equipped with a rapid SSD drive can promote data read and write speeds.

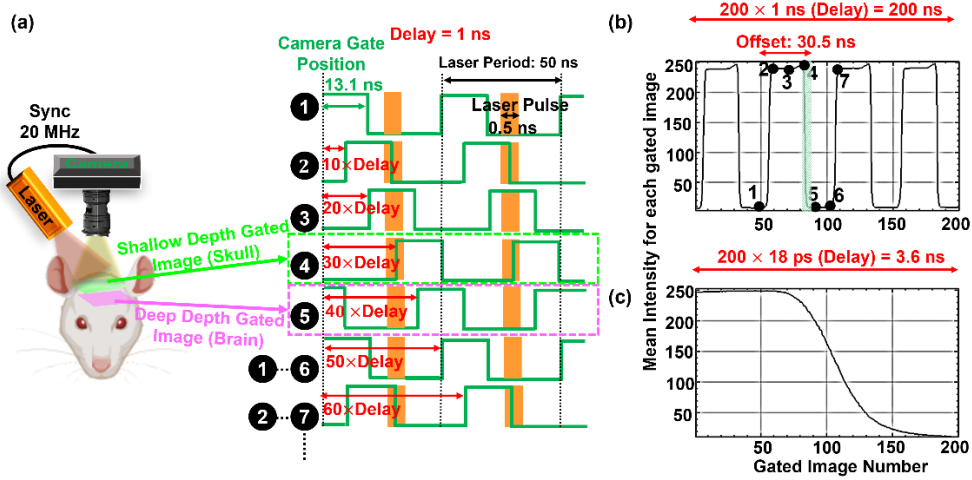

**Figure 2: TR-LSCI principle and data acquisition.** (a) Conceptual representation of a TR-LSCI system to capture 2D flow maps at different depths using different gates. For clarity, only 7 gate positions within a time interval of 10 ns (10 × 1 ns delay time) are illustrated. With the selected offset time of 30.5 ns, shallow and deep depth images were captured at gate positions 4 and 5, respectively. (b) Gate camera intensity profile detected from a calibration tissue phantom at different depths (gates). The preset parameters include laser frequency: 20 MHz (50 ns); offset time: 0 ns; gate delay time: 1 ns; number of gates: 200; detection time window: 200 ns = 200 × 1 ns. Generally, the offset time ranging from 30 to 32 ns was chosen from the shaded region between positions 4 and 5. Longer offset time corresponds to deeper imaging depth. (c) Gate camera intensity profile detected from experimental tissue phantoms and animals at different depths (gates) using preset parameters including laser frequency: 20 MHz (50 ns); offset time: 30.5 ns; gate delay time: 18 ps; number of gates: 200; detection time window: 3.6 ns = 200 × 18 ps. Rat head source: Biorender, www.biorender.com with the publication license.

## 2.3 TR-LSCI Data Processing

**Fig. 3** illustrates multiple steps using our new algorithms with parallel computation and convolution functions in MATLAB to rapidly process TR-LSCI data and generate 2D flow maps [61]. Specifically, the SwissSPAD2 camera captures diffuse laser speckle fluctuations resulting from motions of red blood cells in target tissue volume, measured at different depths with different gates. The resulting gated images at different depths numbered from $G_1$ to $G_M$ are stored sequentially to a folder in the computer. The continuously collected time-course gated images are stored in different folders, numbered from $T_1$ to $T_N$. Two nested for-loops are used to process the gated and time-course images, respectively. The outer for-loop (solid green box) processes the gated images ($G_m$) sequentially. By contrast, the inner parfor-loop (solid blue box) from MATLAB Parallel Computing Toolbox processes multiple time-course images ($T_n$) simultaneously, which significantly reduces the computation time.

The dashed red box in **Fig. 3** shows our new fast LSCI method that uses 2D convolution functions from MATLAB Image Processing Toolbox to convert an intensity image obtained with intensity or gated mode to a 2D flow map. Specifically, a pile-up correction first applies on the intensity image as a pre-processing step to correct possible deformations of the decay shape in the timing and intensity of photon events. The following correction formula was used to adjust for pile-up effect: $I_{cor} = log\left(1 - \frac{I_{rec}}{I_{max}}\right)$, where $I_{max}$ is the maximum photon count (1024 for a 10-bit image), $I_{rec}$ is the recorded photon count, and $I_{cor}$ is the corrected photon count [62].

After pile-up correction, the corrected gated image was converted to a speckle contrast image based on LSCI analysis: $K_s = \frac{\sigma_s}{<I>} = \frac{\sqrt{<I^2> - <I>^2}}{<I>}$, where $K_s$ is defined as the ratio of the standard deviation to mean intensity in a pixel window (w = $N_{pixels} \times N_{pixels}$). A pixel window size of $3 \times 3$, $5 \times 5$, or $7 \times 7$ is usually used to balance the detection sensitivity and spatial resolution [15]. In this paper, based on our experimental data, $K_s$ was quantified over an optimal window of 9 pixels (i.e., $N_{pixels} = 3$).

Conventional methods for LSCI analysis use two nested for-loops to iteratively compute $K_s$ values through all pixels, which is time consuming. By contrast, our new LSCI algorithm performs the conv2 function in MATLAB [63] with different kernels to obtain 2D matrixes of $<I>$, $<I^2>$, and $<I>^2$, without the need of time-consuming for-loops. A 2D matrix of $K_s$ is then generated from these convolutions. Multiple time-course $K_s$ maps over a certain period (e.g., $T_N$) are averaged to generate one $K_a$ map with improved signal-to-noise ratio (SNR). Although the exact relationship between the $K_s$ ($K_a$) and flow is nonlinear, blood flow index (BFI) can be approximated as the inverse square of the speckle contrast in LSCI analysis ($BFI \sim \frac{1}{K_s^2}$) [64, 65].

The final step was the correction of hot pixel spots in the reconstructed BFI map. Because of flaws in the manufacturing process, a small fraction of the camera pixels exhibits elevated dark counts (typically 10 or 100 times higher than the median dark count rate), corresponding to the so-called "hot pixels" [62, 66]. The hot pixels are distributed randomly across the camera sensor array, appearing as high-intensity dots (artifacts) on the collected image [59]. Initially, our approach sought to mitigate the hot pixels from raw intensity images prior to the BFI reconstruction process. However, this preliminary endeavor was only partially successful, with traces of the hot pixels persisting despite the correction. In addition, this preprocessing approach incurred significant computational overhead. Consequently, we opted for a post-reconstruction strategy, wherein the hot pixel locations were identified and corrected from the reconstructed BFI map. To fill in the missing values corresponding to the hot pixels, we first tagged their locations and applied iteratively a median filter only on each tagged pixel of the reconstructed BFI map. In this way, the median filter did not affect normal pixels as they were masked out. Experimentally, we found that five iterations are usually enough to correct the hot pixels.

The final outputs from the outer for-loop and inner parfor-loop are gated BFI maps (M). The implementation of parallel computation using parfor-loop (solid blue box) and matrix convolutions in our new LSCI analysis (dashed red box) significantly shortened the time for reconstructing BFI maps. In this study, MATLAB (R2021a) codes run on a desktop computer equipped with the Intel(R) Xeon (R) W-2245 CPU (8 Cores) operating at a frequency of 3.90 GHz. MATLAB uses 8 CPU cores as 8 co-workers for executing parallel computations. When analyzing mouse data consisting of 400 folders ($T_{400}$) with 80 gated images ($G_{80}$) in each folder, the total time required to reconstruct 80 gated BFI maps (including the averaging over 400 folders) was reduced significantly from ~68 hours to 115 seconds. More specifically, the time to reconstruct one gated BFI map was reduced from ~50 minutes to ~5 seconds.

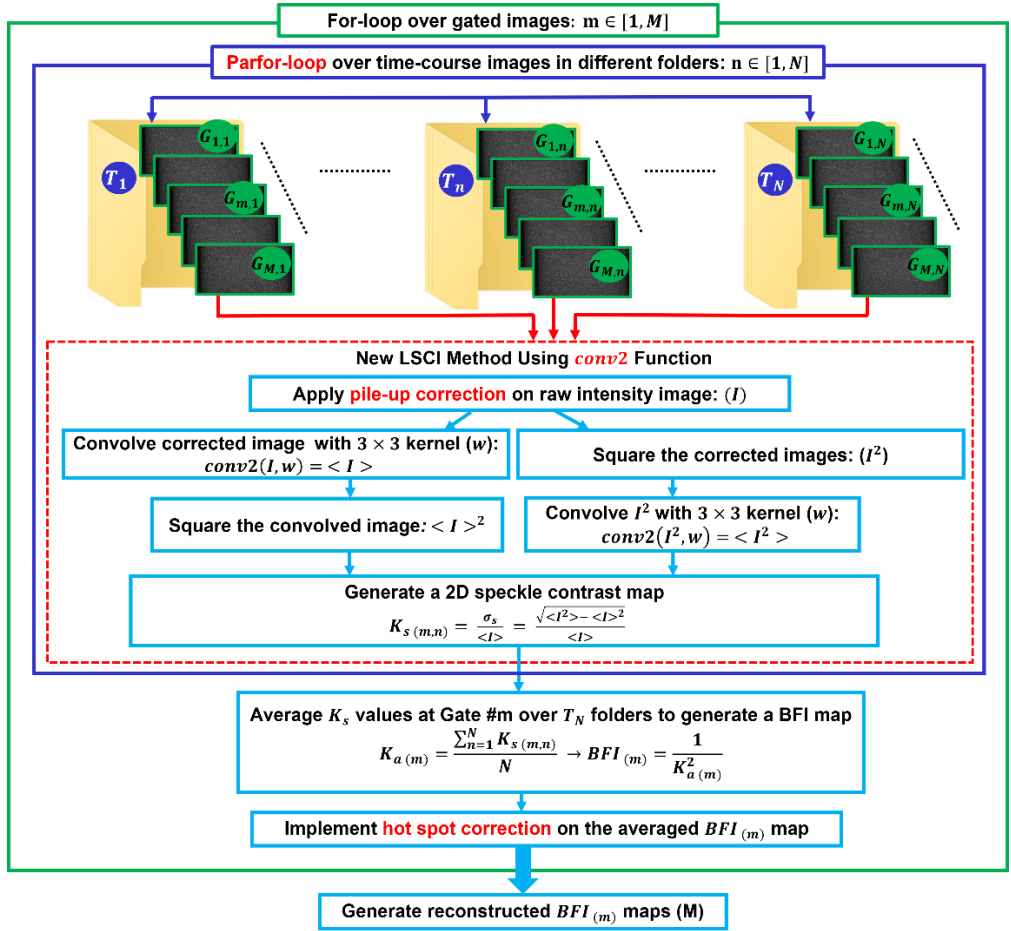

**Figure 3: The flowchart for processing TR-LSCI data and generating 2D CBF maps.** Two nested for-loops were used to carry out computational processes including the outer for-loop to iterate through all gated images (solid green box) and inner parfor-loop to process time-course data in parallel (solid blue box). Steps inside the dashed red box were applied to every intensity gated image to eventually generate a 2D map of $k_s$. **m**: number of gated intensity images; **n**: number of time-course intensity images; $G_{m,n}$: gated image taken at Gate #**m** and Time #**n**; $T_n$: time folder; **w**: pixel window size, $I$: raw intensity image, $<I>$: mean intensity image; $<I>^2$: square of mean intensity image; $<I^2>$: mean of squared intensity image; $k_s$: laser speckle contrast map; $k_a$: average laser speckle contrast at Gate #**m** over all time folders ($T_1:T_N$); **BFI**: blood flow index.

### 2.4 Depth Sensitivity Validation of TR-LSCI Using Head-Simulating Phantoms

Head-simulating phantoms with known optical properties and geometries (**Fig. 4a-4c**) were fabricated and used to illustrate the fundamental concept and depth sensitivity of TR-LSCI. The solid phantoms have a top layer with varied thickness to mimic the skulls of small animals (rodents). Three solid phantoms (no flow) with the empty channels bearing the University of Kentucky logo were fabricated using a 3-D printer (SL1, Prusa) with top layer thicknesses of 1, 2, and 3 mm, respectively. The solid phantoms were made of Titanium dioxide ($TiO_2$), India ink (Black India, MA), and clear resin (eSUN Hard-Tough). The empty UK logo channels were then filled with liquid solutions composed of Intralipid particles (Fresenius Kabi, Sweden), India ink, and water. India ink concentration in the phantoms regulates the absorption coefficient $\mu_a$ while $TiO_2$ and Intralipid concentrations regulate the reduced scattering coefficient $\mu_s'$. Intralipid particles in the liquid phantom provide Brownian motions (i.e., particle flow) to mimic movements of red blood cells in the brain [8, 67]. Optical properties of both solid and liquid phantoms were set in the range of realistic tissue: $\mu_a = 0.03\ cm^{-1}$, and $\mu_s' = 9\ cm^{-1}$ [68, 69].

The noncontact TR-LSCI system was configured to image the ROI of $30 \times 60$ mm$^2$ on the UK logo phantom surface. TR-LSCI collected 200 gated images at different depths sequentially with an interval delay time of 18 ps. Flow contrasts of UK logo phantoms at varied gates/depths were analyzed to demonstrate the depth sensitivity and spatial resolution of TR-LSCI (**Fig. 4d-4g**).

*2.5 In Vivo Study Protocols*

All animal experimental procedures were approved by the UK Institutional Animal Care and Use Committee (IACUC). One mouse and six rats were imaged to assess the performance of TR-LSCI for mapping BFI at different depths. An adult male, C57BL6 mouse (9 months old) was anesthetized with Isoflurane and imaged with both gated and intensity modes to compare imaging spatial resolutions (vessel visualization) as the mouse has a thinner skull than the rat. Six rats were imaged with the gated mode to show the capability of TR-LSCI for 2D mapping of BFI in deeper brain through a thicker skull. Both the mouse and rats were subjected to acquisition of 80 gated images using a constant gate delay time of 18 ps. Eighty gated images (instead of 200) were taken in animal studies to achieve a sampling rate higher than for the phantom experiments. The relative time-course changes in CBF (rCBF) were calculated by normalizing BFI data to the baseline value before the hypercapnia stimuli or cerebral ischemic challenge.

**Baseline Imaging of BFI in a Mouse (Fig. 5).** The mouse on the heating blanket was anesthetized (1-2% isoflurane) with its head secured on a stereotaxic frame. The mouse scalp was surgically removed to reduce its partial volume effect on the deep brain. The noncontact TR-LSCI was positioned above the mouse head and scanned over a ROI of $20 \times 10$ mm$^2$ on the exposed intact skull.

**Continuous Imaging of Global rCBF Increases in Rats during $CO_2$ Inhalations (Fig. 6).** TR-LSCI imaging was continuously performed in six adult male Sprague-Dawley rats (2-3 months) before, during, and after the exposure to a gas mixture consisting of 8%$CO_2$ and 92%$O_2$. It is well known that $CO_2$ is a vascular dilator, leading to a global increase in CBF. The rat on the heating blanket was anesthetized (1-2% isoflurane) with its head secured on a stereotaxic frame. The rat scalp was surgically removed to expose its skull. The noncontact TR-LSCI scanned over a ROI of $30 \times 15$ mm$^2$ on the exposed intact skull for continuous BFI mapping. After a baseline measurement for ~5 minutes, the mixed gas of 8%$CO_2$/92%$O_2$ was administered through a nose cone connected to Matheson Mixer Rota-meter for ~5 minutes. The $CO_2$ was then stopped, and TR-LSCI measurements lasted for ~10 minutes to record BFI recovery towards baseline.

**Continuous Imaging of Regional rCBF Decreases in Rats during Transient Carotid Artery Ligations (Fig. 7).** After the $CO_2$-inhalation experiments, the six rats underwent transient bilateral ligations of common carotid artery (CCA) to create sequential decreases in CBF within the left and right hemispheres. The hairs at cervical surgical site for CCA ligations were shaved and removed with hair cream, and cervical skin was disinfected with Betadine followed by 70% Ethanol. A midline incision was performed to expose and isolate both the left and right CCAs. A sterile surgical suture was wrapped around each CCA, and a loose knot was tied on each suture without impeding blood flow. TR-LSCI configuration and continuous measurements were identical to these described in the $CO_2$ inhalation protocol. After a baseline measurement for ~5 minutes, the right CCA knot was tightened for ~5 minutes to occlude the right CCA. Then the left knot was tightened for ~2 minutes to induce a transient global ischemia. The left and right knots were then released sequentially for ~5 minutes each, allowing the restoration of CBF to the brain. At the end, the animal was euthanized with 100% $CO_2$ inhalation.

*2.6 Statistical Analysis*

Statistical analyses in animal studies were conducted using SPSS software (version 29). Differences in rCBF variations across different phases of stimuli ($CO_2$ inhalations and transient arterial occlusions) were evaluated using repeated measures analysis of variance (ANOVA). A p-value < 0.05 is considered significant for statistical analyses.

## 3. Results

*3.1 TR-LSCI Enables Detection of Flow Contrasts with Depth Sensitivity*

**Fig. 4d-4g** shows resulting 2D maps of particle flow contrasts captured at the gates of #60, #80, #100, and #120 from three UK logo phantoms with the top layer thicknesses of 1, 2 and 3 mm, respectively. One hundred (100) time-course images were taken and averaged at each gate to increase SNRs. Higher flow contrasts were observed at deeper depths with larger gate numbers, indicating the depth sensitivity of imaging by the TR-LSCI. SNRs decreased with the increase of imaging depth (i.e., from Gate #60 to Gate #120) and top layer thickness (i.e., from 1 to 3 mm). These results are expected as deeper penetration and thicker top layer resulted in fewer diffused photons being detected, thus leading to lower SNRs. Results obtained from phantoms indicate that each gate delay of 18 ps corresponds to a 63 µm depth increment.

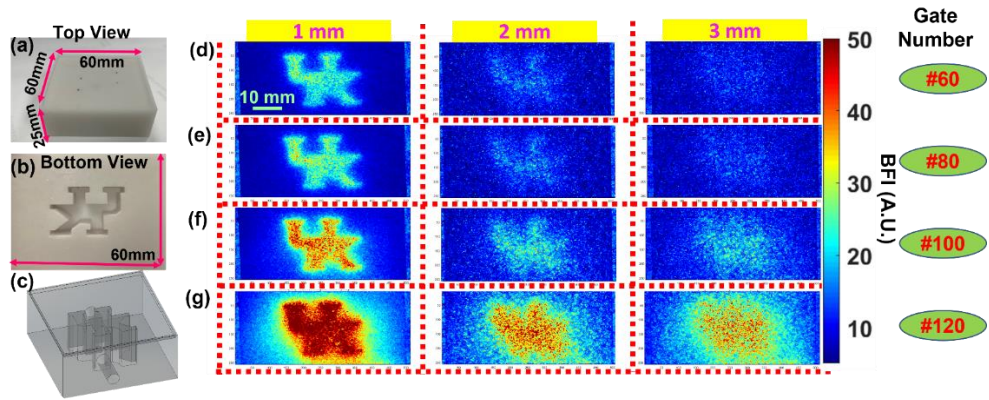

**Figure 4: Experimental results in head-simulating phantoms. (a)-(c)** Top, bottom, and 3D views of a 3D-printed UK logo solid phantom without flow (i.e., flow index = 0) with varying top layer thicknesses (1, 2, 3 mm). The empty UK letter channels were filled in liquid phantom solution with Intralipid particle flow (i.e., flow index = 1) to generate flow contrasts. **(d)-(g)** Reconstructed 2D maps of particle flow contrasts in three phantoms with the top layer thicknesses of 1, 2 and 3 mm respectively, generated by the TR-LSCI at Gates #60, #80, #100, and #120 respectively. The preset parameters for TR-LSCI measurements include laser frequency: 20 MHz (50 ns); offset time: 30.5 ns; gate delay time: 18 ps; number of gates: 200; detection time window: 3.6 ns = 200 × 18 ps. One hundred (100) time-course images at each gate were taken and averaged to increase SNRs.

*3.2 TR-LSCI Gated and Intensity Modes Generate Similar BFI Maps in the Mouse*

**Fig. 5a** shows a raw intensity image of the mouse skull with its scalp retracted. **Fig. 5b** shows the reconstructed BFI map with the gated mode at Gate #0. Gate #0 was selected for the comparison with the intensity mode sensitive to tissue surface. **Fig. 5c** shows the reconstructed BFI map with the intensity mode. The camera exposure time for the intensity mode was 5.12 ms. Four hundred (400) time-course images were taken and averaged for both gated and intensity modes to increase SNRs. The gated and intensity modes generated similar BFI maps.

However, the gated image (**Fig. 5b**) provided more vasculature details than the intensity image (**Fig. 5c**), which is likely due to depth sensitivity of the gated mode.

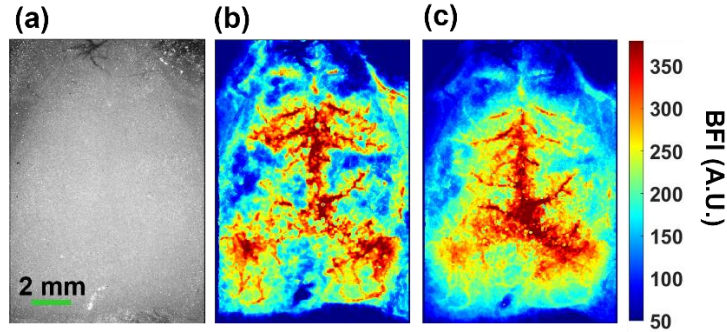

**Figure 5: 2D mapping of BFI during rest status in a mouse.** (**a**) A raw intensity image of mouse skull with its scalp retracted. A zoom lens connected to the TR-LSCI was adjusted to scan a ROI of $20 \times 10$ mm$^2$ on the mouse skull. (**b**) and (**c**) Reconstructed BFI maps with the gated mode at Gate #0 and the intensity mode, respectively. The camera exposure time for the intensity mode was 5.12 ms. Preset parameters for the gated mode include laser frequency: 20 MHz (50 ns); offset time: 30.5 ns; gate delay time: 18 ps; number of gates: 80; detection time window:1.4 ns = $80 \times 18$ ps. Four hundred time-course images were taken and averaged for both intensity and gated modes to increase SNRs.

### 3.3 TR-LSCI Captures rCBF Increases During CO$_2$ Inhalations in Rats

**Fig. 6a** shows a raw intensity image of the rat skull with its scalp retracted. **Fig. 6b** shows BFI maps before, during, and after CO$_2$ inhalation, taken by the TR-LSCI at Gate #35 from an illustrative rat (Rat #4). Gate #35 was selected to ensure sufficient penetration depth to image the rat brain. Based on rat head anatomy, a penetration depth of 2 mm reaches the rat brain cortex through its skull (thickness of ~1 mm) [70]. Since a gate delay of 18 ps is equivalent to ~63 µm depth traveling inside biological tissues [71], imaging depth at Gate #35 exceeds 2 mm (i.e., $35 \times 63$ µm). Approximately 70 time-course images were averaged during the baseline and CO$_2$ inhalation phases respectively to generate BFI maps with improved SNRs. Approximately 140 images were averaged during the recovery phase. **Fig. 6c** shows time-course rCBF changes in the selected ROI in the same illustrative rat (Rat #4), measured continuously by the TR-LSCI with a sampling rate of 0.2 Hz. rCBF changes were calculated by normalizing BFI data to the baseline values prior to CO$_2$ inhalation. **Fig. 6d** shows average time-course changes in rCBF over 6 rats. Average time-course changes in rCBF and p-values for comparing those changes at different phases of CO$_2$ inhalation are summarized in **Table 1** and **Table 2**, respectively. The inhalation of 8%CO$_2$ resulted in a significant increase in rCBF at the endpoint of inhalation compared to the baseline (mean ± standard error: 112.3% ± 3.8%; repeated measures ANOVA: p = 0.026), which is consistent with previous studies [8]. A significant difference (repeated measures ANOVA: p = 0.042) was also observed between the 8%CO$_2$ inhalation and the endpoint of the recovery phase. **Fig. 6e** shows average time-course rCBF changes taken at different gates (Gate#1, Gate #10, Gate #20, Gate #35, Gate #50) over 6 rats. No significant differences in rCBF were found among different gates.

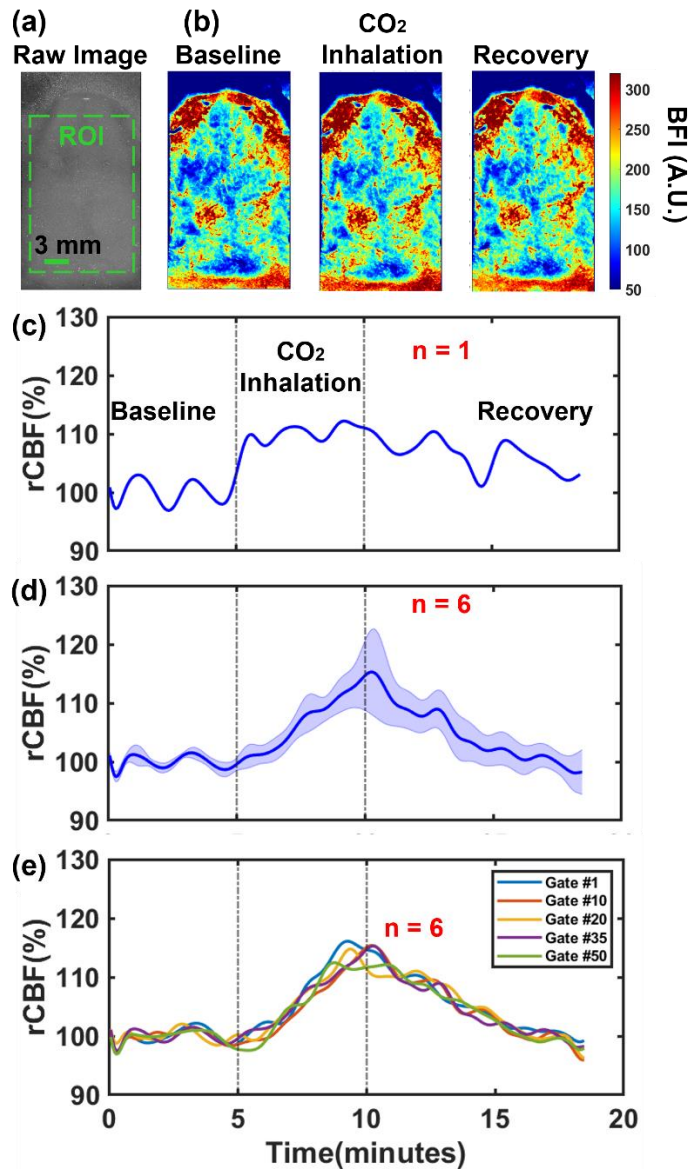

**Figure 6: TR-LSCI imaging of rCBF responses to hypocapnia in 6 rats.** (**a**) A raw intensity image of illustrative rat (Rat #4) skull with its scalp retracted. The selected brain region (ROI) for data analysis is indicated by a dashed green box. (**b**) Reconstructed BFI maps before, during, and after $CO_2$ inhalation, taken at Gate #35. During the baseline and $CO_2$ inhalation phases, around 70 time-course images were averaged to create BFI maps with enhanced SNRs. Similarly, approximately 140 images were averaged during the recovery phase. (**c**) Corresponding time-course rCBF changes in the selected brain region in the same illustrative rat. (**d**) Average time-course rCBF changes (mean value ± standard error) over 6 rats. The shaded area represents standard errors. (**e**) Comparison of average time-course changes in rCBF (mean values) at Gates #1, #10, #20, #35, and #50 over 6 rats. For clarity, only mean values at different gates (i.e., no standard errors) are presented. Preset parameters for the gated mode include laser frequency: 20 MHz (50 ns); offset time: 30.5 ns; gate delay time: 18 ps; number of gates: 80; detection time window: 1.4 ns = 200 × 18 ps.

Table 1. Average rCBF changes (mean ± standard error) from their baselines (100%) during $CO_2$ inhalation over 6 rats

|  | 8%$CO_2$ inhalation | Recovery |
|---|---|---|
| Brain Region | 112.3% ± 3.8% | 99.5% ± 2.3% |

Table 2. Repeated measures ANOVA to assess differences in rCBF changes between different stages of $CO_2$ inhalation

|  |  | p-value |
|---|---|---|
| **Baseline** | 8%$CO_2$ Inhalation | 0.026* |
|  | Recovery | 0.852 |
| **8%$CO_2$ Inhalation** | Baseline | 0.026* |
|  | Recovery | 0.042* |
| **Recovery** | Baseline | 0.852 |
|  | 8%$CO_2$ Inhalation | 0.042* |

*The difference between mean values is significant: $p < 0.05$.

### 3.4 TR-LSCI Captures rCBF Decreases During Transient Artery Ligations in Rats

**Fig. 7a** shows a raw intensity image of the rat skull which was used to generate BFI map. **Fig. 7b** shows BFI maps before, during, and after transient global ischemia, induced by unilateral and bilateral CCA ligations, and during 100%$CO_2$ euthanasia on an illustrative rat (Rat #1). BFI maps were taken by the TR-LSCI at Gate #35. Approximately 70 images were averaged for almost all phases of ligations to improve SNRs, except the transient bilateral ligation period where 28 images were averaged. **Fig. 7c** shows time-course rCBF changes in the same illustrative rat (Rat #1) at the left and right hemispheres, measured continuously by TR-LSCI with a sampling rate of 0.2 Hz. rCBF changes were calculated by normalizing BFI data to the baseline values prior to CCA ligations. **Fig. 7d** shows average time-course rCBF changes at two hemispheres over 4-6 rats. The number of rats for averaging varied due to the unfortunate death of two rats towards to the later phases of experiments. **Table 3** and **Table 4** provide a summary of the average changes in rCBF and p-values for comparing those changes during various phases of CCA ligations. As expected, sequential ligations of the right and left CCAs, followed by their releases, led to significant changes in rCBF in corresponding hemispheres compared to their baselines (repeated measures ANOVA: $p < 0.05$). No significant differences in rCBF were observed among different gates (data are not shown).

Additionally, the euthanasia phase was examined to include all available data (n = 4). Paired t-test was employed to evaluate the difference in rCBF between the baseline and euthanasia phase. Apparently, euthanasia resulted in further reduction in rCBF. During 100%$CO_2$ euthanasia, rCBF values were reduced significantly to 37.6% ± 10.4% (p = 0.009) and 43.1% ± 15.3% (p = 0.034) in right and left hemispheres, respectively (**Table 3**). Notably, rCBF was reduced to the minimal values at the endpoint of 100%$CO_2$ euthanasia, with the values of 29% ± 15.6% and 33% ± 19.3% in the right and left hemispheres, respectively. During the 100%$CO_2$ procedure, one of the four rats exhibited a remarkable resistance to the inhalation and did not succumb to euthanasia. As a result, this rat did not show an obvious decrease in rCBF, leading to a bias in the overall estimation of group rCBF reduction.

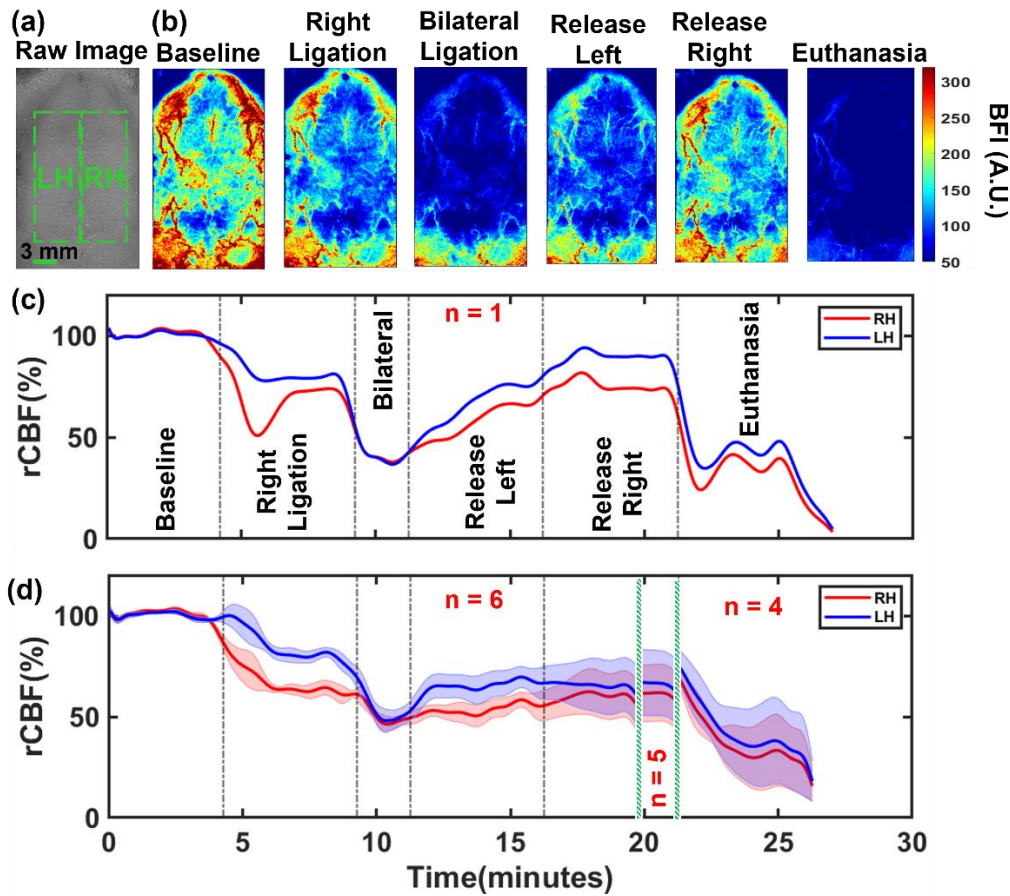

**Figure 7: TR-LSCI imaging of rCBF responses to transient artery ligations in 6 rats.** (a) A raw intensity image of illustrative rat (Rat #1) skull with its scalp retracted. The selected brain regions for data analysis are represented by dashed green boxes on the right hemisphere (RH) and left hemisphere (LH), respectively. (b) Reconstructed BFI maps before, during, and after unilateral and bilateral CCA ligations and releases, followed by $CO_2$ euthanasia taken at Gate #35 in one illustrative rat (Rat #1). For most stages of ligations, ~70 time-course images were averaged to enhance SNRs, while 28 time-course images were averaged during the bilateral ligation phase. (c) Corresponding time-course rCBF changes at two hemispheres (RH and LH) in the same illustrative rat, measured continuously by TR-LSCI with a sampling rate of 0.2 Hz. (d) Average time-course rCBF changes (mean value ± standard error) in each hemisphere over 4-6 rats. The shaded area represents standard errors. Preset parameters for the gated mode include laser frequency: 20 MHz (50 ns); offset time: 30.5 ns; gate delay time: 18 ps; number of gates: 80; detection time window: 1.4 ns = 200 × 18 ps.

**Table 3. Average rCBF changes (mean ± standard error) from their baselines (100%) during unilateral and bilateral CCA ligations over 4 to 6 rats**

|  | Right Ligation n = 6 | Bilateral Ligation n = 6 | Release Left Ligation n = 6 | Release Right Ligation (Recovery) n = 6 | Euthanasia n = 4 |
|---|---|---|---|---|---|
| Right Hemisphere (RH) | 66.7% ± 1.8% | 51.1% ± 2.4% | 53.4% ± 3.9% | 59.7% ± 10% | 37.6% ± 10.4% |
| Left Hemisphere (LH) | 84.4% ± 3.1% | 53.8% ± 4.3% | 64.7% ± 5.5% | 65.6% ± 11.7% | 43.1% ± 15.3% |

Table 4. Repeated measures ANOVA to assess differences in rCBF changes between different stages of ligation in right hemisphere (RH) and left hemisphere (LH)

|  |  | Right Hemisphere (RH) p-value | Left Hemisphere (LH) p-value |
|---|---|---|---|
| **Baseline** | Right Ligation | <0.001* | <0.001* |
|  | Bilateral Ligation | <0.001* | <0.001* |
|  | Release Left Ligation | <0.001* | <0.001* |
|  | Release Right Ligation (Recovery) | 0.004* | 0.010* |
| **Right Ligation** | Baseline | <0.001* | <0.001* |
|  | Bilateral Ligation | <0.001* | <0.001* |
|  | Release Left Ligation | 0.028* | 0.004* |
|  | Release Right Ligation Recovery) | 0.569 | 0.145 |
| **Bilateral Ligation** | Baseline | <0.001* | <0.001* |
|  | Right Ligation | <0.001* | <0.001* |
|  | Release Left Ligation | 0.650 | 0.046* |
|  | Release Right Ligation (Recovery) | 0.462 | 0.319 |
| **Release Left Ligation** | Baseline | <0.001* | <0.001* |
|  | Right Ligation | 0.028* | 0.004* |
|  | Bilateral Ligation | 0.650 | 0.046* |
|  | Release Right Ligation (Recovery) | 0.438 | 0.914 |
| **Release Right Ligation (Recovery)** | Baseline | 0.004* | 0.010* |
|  | Right Ligation | 0.569 | 0.145 |
|  | Bilateral Ligation | 0.462 | 0.319 |
|  | Release Left Ligation | 0.438 | 0.914 |

*The difference between mean values is significant: $p < 0.05$

## 4. Discussion and Conclusions

Continuous and longitudinal monitoring of CBF holds significance for both neuroscience research and clinical applications. In contrast to large, expensive, and single-shot neuroimaging modalities such as MRI and CT, portable and affordable optical imaging instruments enable continuous and longitudinal assessment of CBF at the bedside of clinics. Conventional continuous-wave systems with coherent laser illumination for CBF measurements include LSCI and DCS; both have limitations. LSCI with a widefield illumination and 2D ordinary camera detection enables fast and high-resolution 2D mapping of superficial CBF (depth less than 1 mm) [72] whereas DCS with point sources and discrete photodetectors enables low-resolution 3D tomography of deeper CBF (depth up to centimeters) [73]. Moreover, collected cerebral signals by LSCI and DCS are inherently influenced by partial volume artifacts from overlayer tissues (scalp and skull). More recently, time-resolved systems such as TR-DCS use pulsed point sources and discrete single-photon avalanche diodes to measure temporal autocorrelation functions for quantifying CBF at different depths (i.e., depth sensitive) [9, 74]. However, most

TR-DCS systems suffer from poor spatial resolution and limited head coverages due to limited numbers of discrete sources and detectors [45].

This paper reports an innovative TR-LSCI technique that enables fast and high-resolution 2D mapping of CBF at different depths over a large ROI on the head [50]. TR-LSCI illuminates picosecond-pulsed, coherent, widefield near-infrared light onto the head and synchronizes a fast, time-gated, single-photon camera (SwissSPAD2) to map CBF distributions at different depths of the brain. Similar to LSCI, integration of widefield illumination and 2D camera detection in TR-LSCI enables fast and high-resolution 2D mapping of CBF. By applying the time-resolved (TR) method uniquely, TR-LSCI discards photons with short pathlengths, which predominantly traverse extracerebral layers such as the scalp and skull. Meanwhile, photons with longer pathlengths are retained, which penetrate deeper and ultimately reach the brain.

To prove the concept, a benchtop TR-LSCI system (**Fig. 1**) was developed, *for the first time,* and optimized iteratively [50]. Adjustable optomechanical holders were utilized to install and align the free space pulsed laser and SwissSPAD2 camera on the same ROI. An articulating baseball stage was employed for easy rotation of the light source. Two diffusers were placed in front of the point source to achieve homogeneous widefield illumination. A pair of polarizers were added across the source and detection paths to reduce the influence of source reflection directly from the tissue surface.

To synchronize the laser and camera, the picosecond pulsed laser as the master device provided the timing reference for the operation of the SwissSPAD2 at 20 MHz (**Fig. 2**). The customized control software (MATLAB) allowed for setting up multiple camera parameters used in each experiment, including initial offset time, gate delay, number of gates, and exposure time. Both the offset time and gate delay played a role in determining imaging depth and sensitivity. By adjusting the number of gates to be collected, the sampling rate was also changed. For example, collecting 4 gates versus 80 gates achieved sampling rates of 1 Hz and 0.2 Hz, respectively. On the other hand, the optimization of laser power was essential to adapt the linearity range of SwissSPAD2 camera and comply with the ANSI standard.

To expedite data processing, the new algorithms with parallel computation and convolution functions in MATLAB were developed for processing TR-LSCI data and generating 2D flow maps rapidly (**Fig. 3**). Conventional LSCI algorithms consider the subdivision of an original image into multiple pixel windows ($N_{pixels} \times N_{pixels}$) [75], which are computationally extensive due to the use of multiple nested for loops. We addressed this bottleneck by leveraging the conv2 function in combination with the parallel processing capabilities of multicore calculations in MATLAB. As a result, the reconstruction time to generate a single gated BFI map was reduced remarkably from 50 minutes to only 5 seconds.

To demonstrate the depth-sensitivity of TR-LSCI, unique head-simulating phantoms were designed and fabricated with the top solid layer representing the skull (zero flow) and underneath layer of UK logo filled with Intralipid particle flow (non-zero flow) representing the brain (**Fig. 4**). The top layer thicknesses of 1-3 mm coincide approximately with the skull thicknesses of mice and rats. Results from the phantom measurements illustrate the capability of TR-LSCI for 2D mapping of flow contrasts in the deep "brain" through the surface "skull".

Our analysis of the data from the phantom with 3 mm top layer revealed the emergence of the "UK logo" shape around the gate number of 95 (**Fig. 4f**). Considering that light travels approximately 6 mm into and out of the top layer, this corresponds to an approximate depth increment of ~63 µm for each gate delay (i.e., 6 mm / 95 gates). This finding aligns with previous depth estimation using SwissSPAD2 for subsurface fluorescence imaging, where a linear relationship depending on tissue reduced scattering coefficient ($\mu_s'$) was found between the photon flight time and penetration depth [71]. With the gate delay of 18 ps, a depth increment of ~360 µm was observed in the tissue phantom with $\mu_s' = 45 \ cm^{-1}$. After adjusting the $\mu_s'$ difference in our tissue phantom ($9 \ cm^{-1}$, five times smaller), each gate delay of 18 ps corresponds to a depth increment of ~72 µm (i.e., 360 µm / 5).

To assess the capability of TR-LSCI in capturing cerebral vasculature, we performed *in vivo* mapping of BFI in a mouse with the thinner skull (in contrast to rats) using both gated and intensity modes (**Fig. 5**). The image close to the surface (Gate #0) shows the vascular network with apparent anatomical landmarks revealing more details in the gated mode, as compared to the intensity mode. Note that TR-LSCI with the intensity mode is equivalent to the conventional LSCI for mapping superficial CBF. Based on the scale of mouse brain microvasculature [76], TR-LSCI achieved a spatial resolution of tens of micrometers on the tissue surface (**Fig. 5**). To assess the capability of TR-LSCI in continuous monitoring of global/regional rCBF changes, we measured a group of adult rats during transient cerebral hypocapnia (via $CO_2$ inhalation for 5 minutes) and transient cerebral ischemia (through CCA ligations) (**Fig. 6** and **Fig. 7**). During 8%$CO_2$ inhalation, rCBF increased significantly from 100% baseline to 112.3% $\pm$ 3.8% (**Table 1** and **Table 2**). The results agree fairly well with our previous study in adult rats utilizing speckle contrast diffuse correlation tomography (scDCT) with 3D reconstruction: from 100% baseline to 119% $\pm$ 8% during 10%$CO_2$ inhalation for 10 minutes [8]. The small discrepancy in rCBF increase between the two studies can likely be attributed to the differences in $CO_2$ concentrations (8% versus 10%) and $CO_2$ inhalation duration (5 minutes vs 10 minutes).

The sequential CCA ligations induced significant decreases in rCBF: from 100% baseline to 51.1% $\pm$ 2.4% and 53.8% $\pm$ 4.3% within the right and left hemispheres, respectively (**Table 3** and **Table 4**). Previously using scDCT, Huang et al found that sequential CCA ligations caused significant rCBF reductions from 100% baseline to 34% $\pm$ 10% and 32% $\pm$ 11% in the right and left hemispheres, respectively [8]. The discrepancy in rCBF reductions between the two studies may be attributed to the difference in unilateral CCA ligation durations (5 minutes vs 10 minutes). As expected, 100%$CO_2$ euthanasia resulted in significant reductions in rCBF at the endpoint: 29% $\pm$ 15.6% and 33% $\pm$ 19.3% of their baselines in the right and left hemispheres, respectively. Also, TR-LSCI achieved a spatial resolution of 1-2 millimeters when mapping the deep brain of adult rats (**Fig. 6** and **Fig. 7**).

No significant differences in rCBF responses were observed among different gates during both $CO_2$ inhalations (**Fig. 6e**) and CCA ligations. An underlying impact factor could be the large gate width (13.1 ns) of the SwissSPAD2 camera used in our TR-LSCI system, leading to the detection overlap of early and late photons [77]. The presence of this phenomenon resulted in the overlap of different sample volumes/depths, leading to a decrease in depth sensitivity. The depth sensitivity can be improved by using a newly developed gated camera, SPAD512S, with an achievable gate width of 6 ns [78]. Another potential future improvement involves utilizing a customized SPAD512S camera with integrated microlenses to increase the effective fill factor from 10% to 50% [52, 54, 79], thus improving the imaging sensitivity by a similar amount (3-4 fold) [54]. Also, the SPAD512S with a new operating system (C++) is expected to lead to a higher sampling rate (up to 15 Hz), compared to the SwissSPAD2 (up to 1 Hz). Increasing the sampling rate opens up the possibility for mapping brain functional connectivity [80-82], which is an area of interest for our future investigation. Furthermore, the inclusion of other picosecond-pulsed lasers operating at different near-infrared wavelengths would enable simultaneous imaging of CBF and $ScO_2$ distributions using NIRS principles [83, 84].

In conclusion, we have assembled, optimized, and evaluated a revolutionary depth-sensitive TR-LSCI technology for continuous, fast, and high-resolution imaging of cerebral hemodynamics. We evaluated the performance of TR-LSCI through experiments conducted on head-simulating phantoms and *in-vivo* studies in adult rodents. These pilot studies demonstrated that TR-LSCI enabled mapping CBF variations at different depths with a sampling rate of up to 1 Hz and varied spatial resolutions from tens of micrometers on tissue surface to 1-2 millimeters in the deep brain. The results are generally in agreement with previous studies utilizing other cerebral monitoring techniques and similar experimental protocols. Future incorporation of advanced SPAD cameras with improved gate widths and fill factors holds promise for enhancing the performance of TR-LSCI. Additionally, integration of different near-infrared wavelengths and higher sampling rates would enable simultaneous imaging of multiple cerebral hemodynamic parameters and the exploration of brain functional connectivity. With further improvement and validation in larger populations against established methods, we anticipate offering a noninvasive, noncontact, fast, high-resolution, portable, and affordable brain imager for fundamental neuroscience research in animals and translational studies involving human subjects. Moreover, TR-LSCI holds the potential to be used for noninvasive assessment of many vascular and cellular diseases associated with abnormal tissue hemodynamics including cancers, wounds/burns, and muscular diseases.

**Acknowledgements**

We acknowledge partial financial support from the National Institutes of Health (NIH) #R01 EB028792, #R01-HD101508, #R21-HD091118, #R21-NS114771, #R41-NS122722, #R42-MH135825, #R56-NS117587 (G.Y.) and the Halcomb Fellowship in Medicine and Engineering at the University of Kentucky (F.F.). This work was also supported, in part, by the Swiss National Science Foundation (grants 20QT21_187716 Qu3D "Quantum 3D Imaging at high speed and high resolution" and 200021_166289). We would like to thank Biorender for providing us with a licensed rat head image, which has been included in Fig. 2.

**Disclosure**

Guoqiang Yu: Bioptics Technology LLC (F,P), Chong Huang: Bioptics Technology LLC (F,P), Lei Chen: Bioptics Technology LLC (F,P), Faraneh Fathi: (P), Siavash Mazdeyasna: (P), Mingjun Zhao: (P).

Guoqiang Yu is currently collaborating with Bioptics Technology LLC to commercialize the presented TR-LSCI device. Edoardo Charbon holds the position of Chief Scientific Officer of Fastree3D, a company making LiDARs for the automotive market, and Claudio Bruschini and Edoardo Charbon are co-founders of Pi Imaging Technology. Neither company has been involved with the work nor the paper drafting.

**Data Availability**

Data underlying the results presented in this paper are not publicly available at this time but may be obtained from the authors upon reasonable request.